\documentclass[times,rnaas]{aastex631}

\usepackage{graphicx}

\usepackage{amsmath}
\usepackage{bm}
\usepackage{textgreek}

\usepackage{CJK}

\received{May 25, 2021}
\accepted{June 2, 2021}

\shorttitle{Solar Polar Fields}
\shortauthors{Sun et al.}

\begin{document}

\begin{CJK*}{UTF8}{gbsn}

\title{Are the Magnetic Fields Radial in the Solar Polar Region?}

\author[0000-0003-4043-616X]{Xudong Sun (孙旭东)}
\affil{Institute for Astronomy, University of Hawai`i at M\={a}noa, Pukalani, HI 96768, USA; \href{mailto:xudongs@hawaii.edu}{xudongs@hawaii.edu}}

\author[0000-0002-0671-689X]{Yang Liu (刘扬)}
\affil{W.W. Hansen Experimental Physics Laboratory, Stanford University, Stanford, CA 94305, USA}

\author[0000-0003-4446-1696]{Ivan Mili{\'c}}
\affil{Department of Physics, University of Colorado, Boulder, CO 80309, USA}
\affil{Laboratory for Atmospheric and Space Physics, University of Colorado, Boulder, CO 80303, USA}
\affil{National Solar Observatory, Boulder, CO 80303, USA}

\author[0000-0002-0570-4029]{Ana Bel{\'e}n Gri{\~n}{\'o}n-Mar{\'i}n}
\affil{W.W. Hansen Experimental Physics Laboratory, Stanford University, Stanford, CA 94305, USA}
\affil{Institute of Theoretical Astrophysics, University of Oslo, N-0315 Oslo, Norway}
\affil{Rosseland Centre for Solar Physics, University of Oslo, N-0315 Oslo, Norway}


\begin{abstract}
We investigate the orientation of the photospheric magnetic fields in the solar polar region using observations from the Helioseismic and Magnetic Imager (HMI). Inside small patches of significant polarization, the inferred magnetic field vectors at $1\arcsec$ scale appear to systematically deviate from the radial direction. Most tilt towards the pole; all are more inclined toward the plane of sky compared to the radial vector. These results, however, depend on the ``filling factor'' $f$ that characterizes the unresolved magnetic structures. The default, uninformative $f\equiv1$ for HMI will incur larger inclination and less radial fields than $f<1$. The observed trend may be a systematic bias inherent to the limited resolution.
\end{abstract}

\section{}
\vspace{-10mm}


The photospheric magnetic fields in the solar polar region (polar fields) are often assumed to be radial. The radial component $B_r$ is then related to the line-of-sight (LOS) component $B_l$ as $B_r = B_l \mu^{-1}$, where $\mu$ is the cosine of the heliocentric angle. Both observation and modeling lend support to the assumption for spatial scales greater than several heliographic degrees \citep{svalgaard1978,wang1992,petrie2009}.

\begin{figure*}[t!]
\centerline{\includegraphics[width=0.90\textwidth]{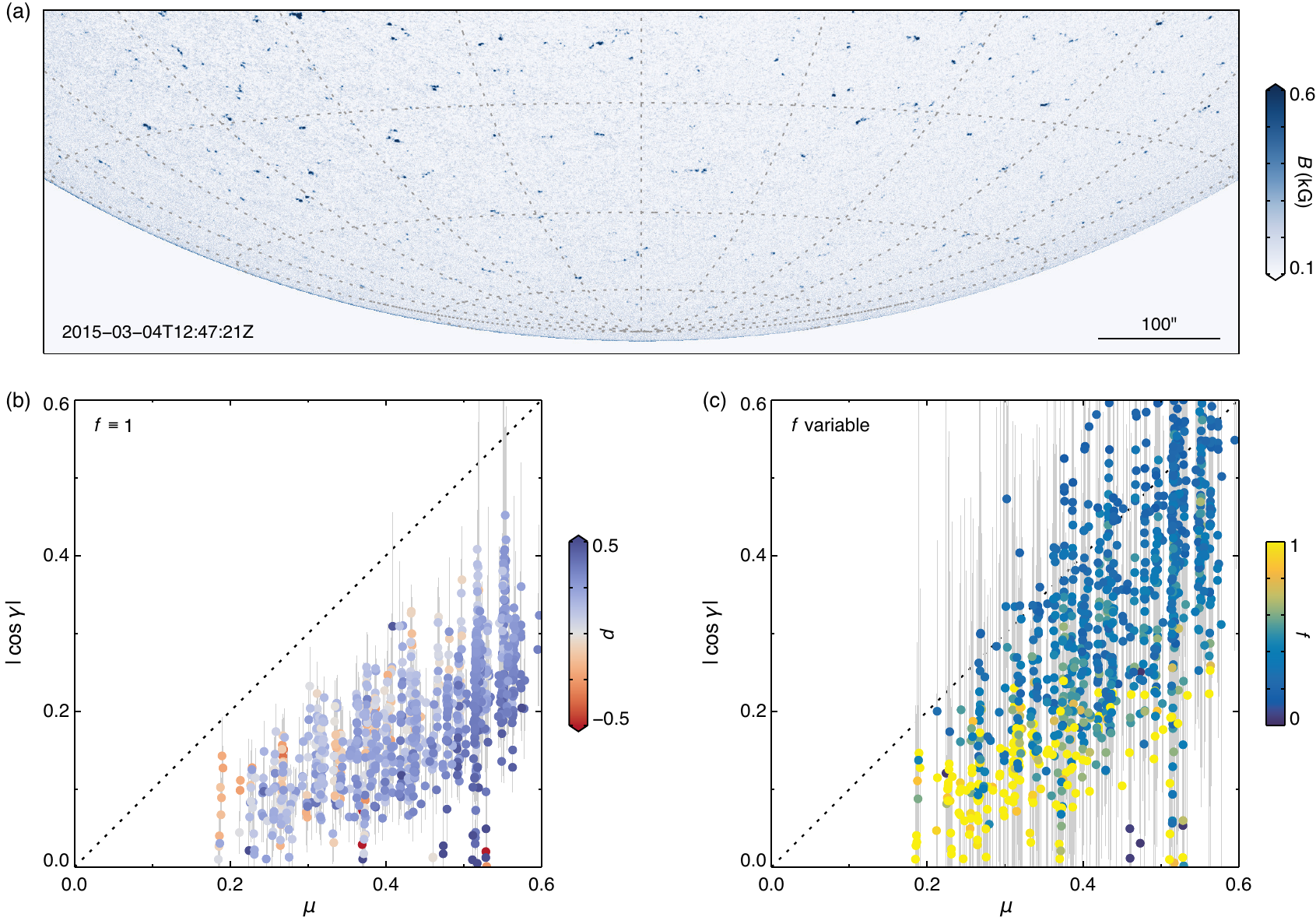}}
\caption{HMI observations of the southern polar region. (a) Map of $B$, inferred with $f\equiv1$. The dotted lines show the latitude (longitude) with $10\degr$ ($15\degr$) spacing.  (b) Scatter plot of $\mu$ vs $\left|\cos \gamma\right|$, inferred with $f\equiv1$. The colors indicate $p=\mathrm{sgn}(B_r)B_\theta B^{-1}$. For the southern hemisphere, $\bm{B}$ tilts toward the pole if $p>0$. The error bars show $\pm\sigma_\gamma \sin \gamma$, where $\sigma_\gamma$ is the formal uncertainty of $\gamma$. (c) Same as (b), but inferred with $f$ as a free parameter. The colors indicate $f$.}
\label{f:fig1}
\end{figure*}

The polar fields are spatially non-uniform. Most polarization signals come from small, isolated patches a few megameter wide, covering at most a few percent of the polar region (Figure~\ref{f:fig1}(a)). The intrinsic field strength $B$ can exceed $1$~kG based on the high-resolution observations ($0\farcs3$) from the \textit{Hinode} satellite \citep{tsuneta2008}.

We investigate the orientation of the magnetic field vectors $\bm{B}$ inside these patches using the Helioseismic and Magnetic Imager (HMI) aboard the \textit{Solar Dynamics Observatory}. HMI provides full-disk Stokes observations of the photospheric \ion{Fe}{1} 617.3~nm line. The spatial resolution is $1\arcsec$ with a $0\farcs5$~pixel$^{-1}$ plate scale; the spectral sampling is performed at six wavelengths. We evaluate the polarization degree for each pixel in the southern polar region measured on March 4, 2015, and select about $1000$ pixels ($0.2\%$ of total) whose signal is above $5\sigma$ of for further analysis. The field of view is limited to $\left[-85\degr,-60\degr\right]$ latitude, and $\left[-75\degr,75\degr\right]$ longitude. The HMI pipeline infers $\bm{B}$ and the formal measurement uncertainties \citep{hoeksema2014}.

The magnetic inclination $\gamma$, defined as $0\degr$ toward and $180\degr$ away from the observer, exhibits an interesting trend. The values of $\left|\cos\gamma\right|$ are invariably smaller than $\mu$ (Figure~\ref{f:fig1}(b)). In other words, the $\bm{B}$ vectors are more inclined toward the plane of sky compared to the local radial vector. About $93\%$ of the pixels also have $\mathrm{sgn}(B_r)B_\theta B^{-1}>0$, where $B_r$ and $B_\theta$ are the radial and meridional components in a heliocentric spherical coordinate. This means that most $\bm{B}$ vectors tilt toward the pole. The median deviation from the radial direction is $18\degr$.

Is this trend genuine, or is it due to some systematic bias? We explore the effect of the ``filling factor'' parameter $f$ in the magnetic inference procedure, which represents the sub-pixel fractional area occupied by the unresolved magnetic structures. In the weak-field approximation \citep{jefferies1991}, $\bm{B}$ depends on the Stokes parameters $(Q,U,V)$ as
\begin{equation}
\begin{split}
f B \cos\gamma & \propto V, \\
f B^2 \sin^2\gamma & \propto ( Q^2 + U^2 )^{1/2}. \\
\end{split}
\label{eq:wkfld}
\end{equation}
Therefore the inferred $\gamma$ depends on $f$ as
\begin{equation}
\tan \gamma \propto f^{1/2} \dfrac{( Q^2 + U^2 )^{1/4}}{V}.
\label{eq:gamma}
\end{equation}

For HMI's moderate resolution, $f$ proves to be highly degenerate with $B$ \citep{centeno2014}. The HMI pipeline thus adopts $f\equiv1$, which is appropriate when no additional information is available. However, the magnetic fields may be structured on scales smaller than the HMI resolution. For the polar magnetic patches, the distribution of $f$ has a broad peak centered around $0.15$ according to \textit{Hinode} observations. If so, Equation~(\ref{eq:gamma}) posits that HMI's unity $f$ will incur a $\left|\tan\gamma\right|$ that is systematically too large, or, a $\bm{B}$ vector too inclined toward the plane of sky.

We test this line of reasoning by re-inferring the HMI data with $f$ as a free parameter (Gri{\~n}{\'o}n-Mar{\'i}n et al., in preparation). The values of $f$ show a broad peak centered at about $0.25$. The field vectors indeed become more radial, in particular for smaller $f$'s, even though the bias persists (Figure~\ref{f:fig1}(c)). The degeneracy between $f$ and $B$ manifests in their correlation coefficients, most of which lie between $-1$ and $-0.8$. Due to the degeneracy, the individual parameter uncertainties increase significantly.

Polar field observations from the Vector Stokes Magnetograph (VSM) instrument appear to have an opposite bias, i.e., the $\bm{B}$ vectors systematically tilt toward the equator \citep{gosain2013}. There is generally $\left|B_r\right| < \left|B_l\right| \mu^{-1}$ for VSM, and $\left|B_r\right| > \left|B_l\right| \mu^{-1}$ for HMI in the polar region. A VSM-HMI discrepancy of the field orientation also shows up in the plage region \citep{pevtsov2021}. The differences are partially credited to the treatment of $f$, in line with our finding.

Previous work shows that the fine-scale fields cannot always be reliably inferred from low-resolution maps \citep{leka2012}. The problem may benefit from high-resolution observations from the \textit{Danial K. Inouye Solar Telescope} \citep{rimmele2020} or out-of-ecliptic observations from the Polarimetric and Helioseismic Imager \citep{solanki2020} aboard the \textit{Solar Orbiter}. They can provide more informative priors for $f$ and $\gamma$, respectively. Synthetic data from numerical simulations will provide a useful benchmark for assessing the systematic biases \citep{plowman2020}.


\begin{acknowledgments}
We thank K. D. Leka for helpful discussions. X. Sun is supported by NASA HGIO award 80NSSC21K0736. The \textit{SDO} data are courtesy of NASA and the HMI science team.
\end{acknowledgments}


\end{CJK*}


\end{document}